\documentclass[conference]{IEEEtran}
\IEEEoverridecommandlockouts

\usepackage[utf8]{inputenc}
\usepackage{cite}
\usepackage{amsmath,amssymb,amsfonts}

\usepackage{algorithm, algorithmic}
\usepackage{graphicx}
\usepackage{textcomp}
\usepackage{xcolor}
\usepackage{booktabs}
\usepackage{multirow}
\usepackage{makecell}
\usepackage{url}
\usepackage{upgreek}

\DeclareRobustCommand*{\IEEEauthorrefmark}[1]{%
    \raisebox{0pt}[0pt][0pt]{\textsuperscript{\footnotesize #1}}%
}

\def\BibTeX{{\rm B\kern-.05em{\sc i\kern-.025em b}\kern-.08em
    T\kern-.1667em\lower.7ex\hbox{E}\kern-.125emX}}
\begin{document}

\title{DX2CT: Diffusion Model for 3D CT Reconstruction from Bi or Mono-planar 2D X-ray(s)}

\author{
\IEEEauthorblockN{Yun Su Jeong\IEEEauthorrefmark{1,$\star$}, Hye Bin Yoo\IEEEauthorrefmark{1,$\star$}, Il Yong Chun\IEEEauthorrefmark{1,2,3,$\dag$}
}
\IEEEauthorblockA{\IEEEauthorrefmark{1}Department of Electrical and Computer Engineering, Sungkyunkwan University, Republic of Korea}
\IEEEauthorblockA{\IEEEauthorrefmark{2}Departments of Artificial Intelligence and Advanced Display Engineering, Sungkyunkwan University, Republic of Korea}
\IEEEauthorblockA{\IEEEauthorrefmark{3}Center for Neuroscience Imaging Research, Institute for Basic Science (IBS), Suwon 16419, Republic of Korea}

\thanks{
    $^\star$Equal contribution.
    $^\dag$Corresponding author.
    Emails of authors: \newline
    \noindent $\{ \textsf{intyeger@g.skku.edu}, \textsf{gpqls7662@g.skku.edu}, \textsf{iychun@skku.edu} \}$.
    The work of Y. S. Jeong, H. B. Yoo and I. Y. Chun was supported in part by NRF Grant RS-2023-00213455 funded by MSIT, and the BK21 FOUR Project. 
    The work of I. Y. Chun was additionally supported in part by 
    IITP Grant RS-2019-II190421 funded by MSIT, 
    KIAT Grant RS-2024-00418086 funded by MOTIE,
    IBS-R015-D2, 
    the KEIT Technology Innovation Program Grant 20014967 funded by MOTIE, and 
    the 2024 Digital Therapeutics Development and Demonstration Support Program funded by MSIT and NIPA.
    }
}

\maketitle

\begin{abstract}
Computational tomography (CT) provides high-resolution medical imaging, but it can expose patients to high radiation.
X-ray scanners have low radiation exposure, but their resolutions are low.
This paper proposes a new conditional diffusion model, DX2CT, that reconstructs three-dimensional (3D) CT volumes from bi or mono-planar X-ray image(s).
Proposed DX2CT consists of two key components: \textit{1)} modulating feature maps extracted from two-dimensional (2D) X-ray(s) with 3D positions of CT volume using a new transformer and \textit{2)} effectively using the modulated 3D position-aware feature maps as conditions of DX2CT.
In particular, the proposed transformer can provide conditions with rich information of a target CT slice to the conditional diffusion model, enabling high-quality CT reconstruction.
Our experiments with the bi or mono-planar X-ray(s) benchmark datasets show that proposed DX2CT outperforms several state-of-the-art methods.
Our codes and model will be available at: \url{https://www.github.com/intyeger/DX2CT}.
\end{abstract}

\begin{IEEEkeywords}
Three-dimensional (3D) reconstruction, Diffusion models, Computed tomography (CT), X-ray radiography.
\end{IEEEkeywords}

\section{Introduction}
\label{sec:intro}

Computational tomography (CT) can provide high-resolution three-dimensional (3D) volumes, but can expose patients to high radiation. 
This could limit its use in pediatric patients or patients with the risk of cancers.
X-rays are safe in the perspective of radiation doses, but their resolutions are lower compared to CT.
Recently, researchers have proposed different deep learning approaches that reconstruct 3D CTs from bi or mono-planar two-dimensional (2D) X-ray image(s) \cite{single_tomo, single_proj, x2ctgan, ccxraynet, cvae-gan, x-ctrsnet, perx2ct, xctdiff, diff2ct, difr3ct}.
These approaches can significantly reduce the radiation exposure.
However, it is extremely challenging to reconstruct accurate 3D CTs, particularly 3D fine details, from a few X-ray image(s).

The early convolutional neural network (CNN) models use a 2D encoder-to-3D decoder structure that reconstructs 3D CTs from a monoplanar X-ray image with several 2D and 3D convolutional layers \cite{single_tomo, single_proj}.
Since it is challenging to capture sufficient depth information using a single X-ray,
the later work X2CT-GAN uses biplanar perpendicular X-ray images to complement depth information from each other \cite{x2ctgan}.
X2CT-GAN uses a generative adversarial network (GAN) \cite{gan} with two 2D encoding-to-3D decoding networks as a generator.
Its subsequent methods are modified versions of X2CT-GAN to improve its performance \cite{ccxraynet,cvae-gan, x-ctrsnet}.
However, because these methods do not consider the positional information of 2D X-ray features in a 3D CT volume,
they struggle to capture accurate internal structures in a 3D CT volume, leading to degraded reconstruction performances.
To overcome this limitation, PerX2CT uses a simple 2D encoder-to-2D decoder structure and a non-learnable module that can form 3D position-aware feature maps by relocating X-ray features with the perspective projection of X-ray \cite{perx2ct}.
Yet, this method does not use a learnable module that can adaptively generate 3D position-aware feature maps from 2D X-ray images.

With the success of diffusion models in generating diverse high-quality images \cite{ddpm,ddim, ldm,sr3,sdm, ddrm, light-field}, 
a couple of conditional diffusion models have been proposed for 3D CT reconstruction from 2D X-ray image(s), based on 3D U-Net \cite{3dunet} with high computation and memory demands \cite{xctdiff,diff2ct,difr3ct}.
They extract 3D structural features from X-ray(s) by a learnable module, and use them as condition via simple concatenation or the standard cross-attention module.

In this paper, we propose a new conditional diffusion model, DX2CT, that reconstructs three 3D CT volumes from 2D X-ray image(s).
For computational efficiency, we reformulate the 2D X-ray(s) to 3D CT reconstruction problem into 2D X-ray(s) to 2D CT slices reconstruction problem.
To reconstruct accurate 3D fine details by compensating insufficient 3D inductive bias caused by the reformulation,
we propose a new learnable Vision Transformer (ViT) \cite{vit} that modulates extracted X-ray feature maps with positional information of 3D CT volume and given X-rays to generate 3D position-aware feature maps.
By effectively incorporating modulated 3D position-aware feature maps that capture 3D internal/structural information of CTs into a 2D U-Net \cite{2dunet} of conditional diffusion model, we can improve the reconstruction qualities.

Our contributions are summarized as follows:
\begin{itemize}
    \item We propose a new conditional diffusion model {\bfseries DX2CT} that reconstructs high-quality 3D CT volumes from bi or mono-planar X-ray(s) using 3D positional information in a CT volume.
    \item We propose a new ViT {\bfseries 3D Positional Query Transformer (3DPQT)} that produces X-ray(s) feature maps aligned with 3D positions of a target slice in 3D CT volume.
    \item We effectively incorporate the obtained 3D position-aware feature maps of X-ray(s) as conditions into a denoising 2D U-Net.
    \item Our numerical experiments with the bi or mono-planar X-ray(s) conducted on benchmark The Lung Image Database Consortium (LIDC) \cite{lidc} datasets show that proposed DX2CT outperforms several state-of-the-art (SOTA) methods.
\end{itemize}

\section{Backgrounds}
\label{sec:background}

\subsection{Diffusion models}

Denoising diffusion probabilistic models (DDPMs) \cite{ddpm, ddim, sr3, ldm, sdm, ddrm, light-field} consist of two major components, forward and reverse processes.
The forward process that gradually adds isotropic Gaussian noise to a sample $\mathbf{x}_0$ is defined as
$\mathbf{x}_t = \sqrt{\alpha_t}\mathbf{x}_0 + \sqrt{1 - \alpha_t}\mathbf{\upepsilon}$, 
$t = 0,\ldots,T$, where 
$\mathbf{x}_t$ and $\alpha_t$ denote corrupted sample and a constant that depends on noise schedule at the $t$th timestep, respectively, 
and $\mathbf{\upepsilon} \sim \mathcal{N}(\mathbf{0}, \mathbf{I})$.
The reverse process predicts noise added at the $t$th timestep in the forward process through a trained denoising U-Net and gradually denoises noisy sample $\mathbf{x}_t$ to $\mathbf{x}_{t-1}$.
Starting from $\mathbf{x}_T \sim \mathcal{N}(\mathbf{0}, \mathbf{I})$, the reverse process is repeated until $\mathbf{x}_0$ is generated.
The reverse process of conditional DDPMs \cite{sr3, ldm, sdm, ddrm, light-field} is affected by additional information so-called condition.
By changing condition(s) and conditioning approach, one can manipulate generation outcomes.

\subsection{Vision transformers}
\label{sec:bg:vit}

ViTs \cite{vit} handle vision tasks using the transformer architecture \cite{transformer} and demonstrate high performance across various vision applications \cite{pvt,dit,detr}.
This success is attributed to the attention mechanism that enables models to focus on relevant information in an input image.
ViTs divide an input image into patches and tokenize them, add positional embeddings to provide positional information to tokens, and use these embedding vectors as query, key and value for attention.
In the attention process, the information in query and value is combined by considering content and positional relationships between the query and key.
The transformer encoder takes position-embedded patches as an input and produces the final features via the above attention process.

\begin{figure}[t!]
    \centering
    \includegraphics[width=0.9\linewidth]{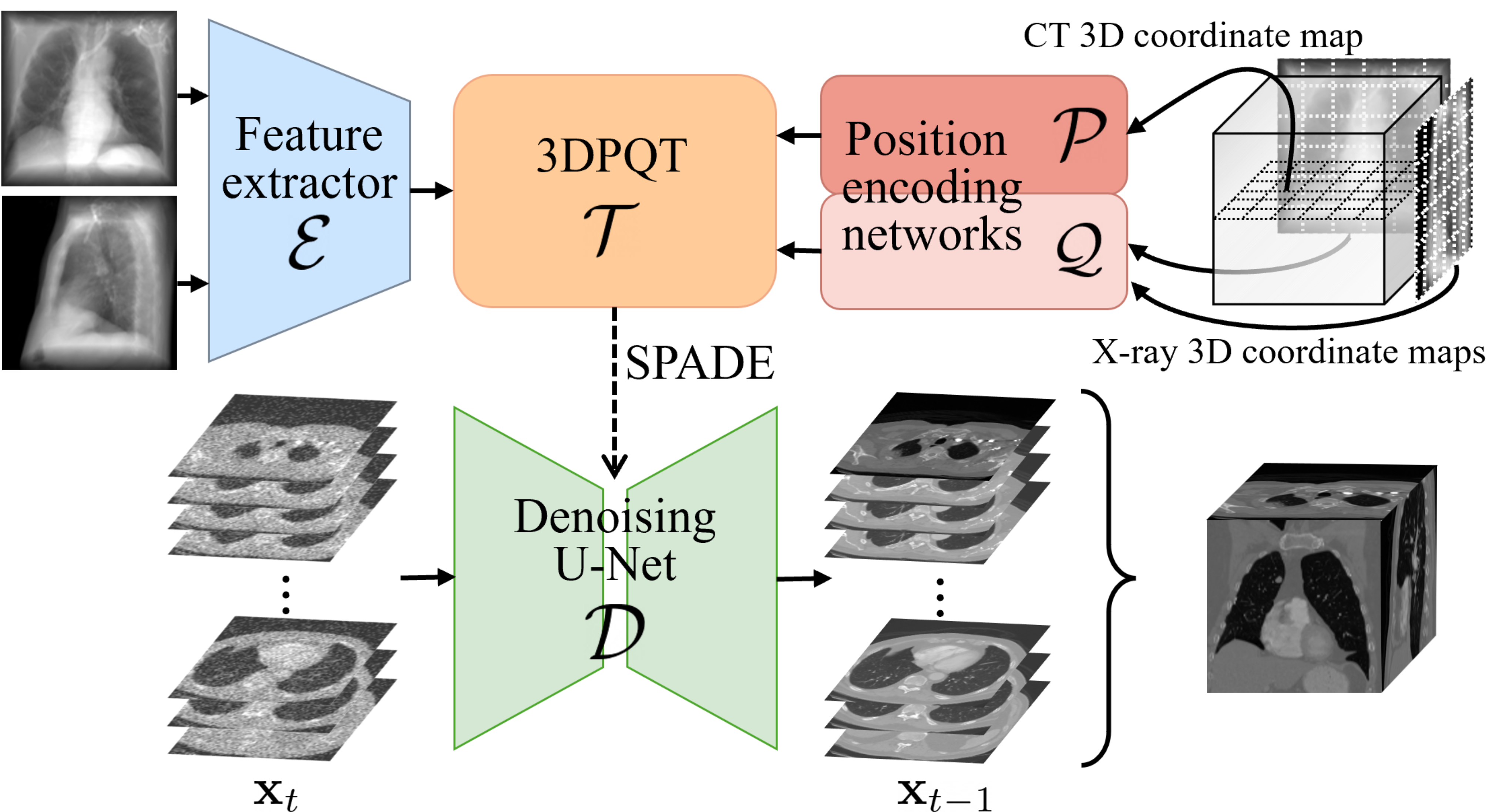}
    \caption{
    Overview of proposed DX2CT in the reverse diffusion process.
    We extract multi-scale feature maps from biplanar X-rays with a feature extractor $\mathcal{E}$ and modulate the extracted feature maps with 3D positional information sets of target CT slice and X-rays that are generated from position-encoding networks $\mathcal{P}$ and $\mathcal{Q}$, respectively.
    We use the spatially-adaptive normalization (SPADE) method to incorporate modulated feature maps into denoising 2D U-Net \cite{2dunet} $\mathcal{D}$ of conditional diffusion model.
    We stack generated CT slices to reconstruct a 3D CT volume.
    We repeat the process for each anatomical plane.
    }
    \label{fig:overview}
\end{figure}

\section{Methodology}
\label{sec:method}

This section proposes the DX2CT framework, where we mainly introduce its default setup that uses perpendicular biplanar X-ray images from posterior-anterior (PA) and lateral (Lat) views.
Figure~\ref{fig:overview} illustrates the overall architecture of DX2CT.
\S\ref{sec:feature map extraction}--\ref{sec: 3DPQT} explain how to extract multi-scale feature maps from X-rays and modulate them by 3DPQT modules.
\S\ref{sec:conditional diffusion model} explains how to incorporate modulated feature maps as conditions into our conditional DDPM model and train DX2CT.

\subsection{Feature map extraction from X-ray images}
\label{sec:feature map extraction}

We extract feature maps from biplanar 2D X-ray images  $\textbf{i}^{\text{PA}}, \textbf{i}^{\text{Lat}} \in \mathbb{R}^{H \times W}$ using the feature extractor $\mathcal{E}_{\theta_\mathcal{E}}$ with parameters $\theta_\mathcal{E}$ and $L$ layers.
We extract multi-scale feature maps with $L$ scales, where feature maps extracted from each layer form those at each scale:
\begin{equation}
    \mathbf{f}^{v} = \{ \mathbf{f}^{v}_{1}, \ldots, \mathbf{f}^{v}_{L} \} = \mathcal{E}_{\theta_{\mathcal{E}}}(\mathbf{i}^{v}),
    \label{eq:x-ray-fe}
\end{equation}
in which $\mathbf{f}^{v}_{l} \in \mathbb{R}^{H_l \times W_l \times C_l}$ denotes extracted feature map from $l$th layer of $\mathcal{E}_{\theta_\mathcal{E}}$,
$l = 1, \ldots, L$,
and
$v \in \left\{ \text{PA}, \text{Lat} \right\}$.

\subsection{Modulating feature maps with proposed 3DPQTs}
\label{sec: 3DPQT}

To generate 3D position-aware feature maps -- corresponding to reshaped final features in ViT (see \S\ref{sec:bg:vit}) -- we propose a new ViT 3DPQT.
The proposed 3DPQT modulates the extracted X-ray feature maps in (\ref{eq:x-ray-fe}) with the 3D positional information of X-rays and target CT slice by learning the positional relationship between 3D CT and two 2D X-rays.

First, to generate CT 3D positional embeddings (PEs) from 3D positions of target CT slices, we propose a position-encoding network $\mathcal{P}_{\theta_{\mathcal{P}}}$ with parameters $\theta_\mathcal{P}$.
We use a 3D Cartesian coordinate system $(x,y,z)$ with $x,y,z \in [-1,1]$ for a target 3D CT volume.
Given an anatomical plane $ m \in \{\text{axial}, \text{coronal}, \text{sagittal}\}$,
we define a map with 3D coordinates of the $n$th slice as $\mathbf{s}^m_n \in \mathbb{R}^{H \times W \times 3}$, $n = 1, \ldots, N$, where $N$ is the total number of slices.
We pass each coordinate in $\mathbf{s}^m_n$ through a $\mathcal{P}_{\theta_{\mathcal{P}}}$ that consists of positional encoding \cite{nerf}, a multi-layer perceptron (MLP) network, and average pooling with appropriate window sizes, resulting in maps of CT PEs.
In particular, by using average pooling with different window sizes,
we generate multi-scale CT PEs that have the same 2D spatial resolution as the multi-scale X-ray feature maps.
The above process can be represented as follows:
\begin{equation}
    \label{eq:pe-ct}
    \mathbf{p}_{n}^{m} = \{ \mathbf{p}_{n,1}^{m}, \ldots, \mathbf{p}_{n,L}^{m} \} = \mathcal{P}_{\theta_{\mathcal{P}}}(\mathbf{s}_{n}^{m}),
\end{equation}
where $\mathbf{p}_{n,l}^{m} \in \mathbb{R}^{H_l \times W_l \times C_\mathcal{P}}$ denotes CT PEs at the $n$th slice and the $l$th scale on an $m$ plane and $C_\mathcal{P} \gg 3$.
In addition, for each X-ray, we modify the above process to generate multi-scale X-ray 3D PEs with either $(x,0,z)$ or $(0,y,z)$ and use another position-encoding network $\mathcal{Q}_{\theta_{\mathcal{Q}}}$ with parameters $\theta_\mathcal{Q}$ to generate multi-scale X-ray PEs,
\begin{equation}
\label{eq:pe-xray}
\mathbf{q}^v = \{ \mathbf{q}^v_{l} \in \mathbb{R}^{H_l \times W_l \times C_l} : l\!=\!1,\dots,L \}, v \in \left\{ \text{PA}, \text{Lat} \right\}.
\end{equation}

Next, we propose the 3DPQT modules $\mathcal{T}_{\theta_{\mathcal{T}}}$ with parameters $\theta_{\mathcal{T}}$ that generate multi-scale 3D position-aware feature maps $\mathbf{c}^m_n$ by using multi-scale CT and X-ray PEs, and multi-scale X-ray feature maps.
We use $L$ 3DPQTs that correspond to the number of feature scales in \S\ref{sec:feature map extraction}.
Each 3DPQT consists of $B$ multi-head cross-attention blocks.
In the $l$th 3DPQT, by adding X-ray PEs $\{ \mathbf{q}^{\text{PA}}_{l}, \mathbf{q}^{\text{Lat}}_{l} \}$ to $\{ \mathbf{f}^{\text{PA}}_{l}, \mathbf{f}^{\text{Lat}}_{l} \}$ in (\ref{eq:x-ray-fe}) with corresponding scale and flattening them, we obtain two X-ray position-aware flattened feature maps $\Tilde{\mathbf{f}}_{l}^{\text{PA}},\Tilde{\mathbf{f}}_{l}^{\text{Lat}} \in \mathbb{R}^{H_l W_l \times C_l}$.
We concatenate them to form $\tilde{\mathbf{f}}_l \in \mathbb{R}^{2 H_l W_l \times C_l}$, and embed $\tilde{\mathbf{f}}_l$ as key and value that have position-aware X-ray features.
Similarly, we flatten $\mathbf{p}^m_{n,l}$ in (\ref{eq:pe-ct}) to obtain $\tilde{\mathbf{p}}^m_{n,l} \in \mathbb{R}^{H_l W_l \times C_\mathcal{P}}$ and embed it as a query that has positional information of a target slice in a 3D CT volume.
As 3D positional query and X-ray feature key and value pass through multi-head cross-attention blocks in 3DPQT, 
the query retrieves features from X-ray feature value by learning positional relationship between a 3D CT volume and two 2D X-rays.
In short, the multi-scale 3D position-aware feature maps produced by modulating $L$ X-rays feature maps using 3DPQTs $\mathcal{T}_{\theta_\mathcal{T}}$ can be represented as follows:
\begin{equation}
    \mathbf{c}_{n}^{m} = \{\mathbf{c}_{n,1,}^{m}, \ldots, \mathbf{c}_{n,L}^{m}\} = \mathcal{T}_{\theta_{\mathcal{T}}}(\mathbf{f}^{\text{PA}}, \mathbf{f}^{\text{Lat}}, \mathbf{p}_{n}^{m}, \mathbf{q}^{\text{PA}}, \mathbf{q}^{\text{Lat}}),
    \label{eq:condition}
\end{equation}
where $\mathbf{c}^m_{n,l} \in \mathbb{R}^{H_l \times W_l \times C_\mathcal{P}}$ denotes modulated 3D position-aware feature maps from $l$th 3DPQT,
and $\{ \mathbf{f}^\text{PA}, \mathbf{f}^\text{Lat} \}$, $
\mathbf{p}_n^m$, and $\{ \mathbf{q}^\text{PA}, \mathbf{q}^\text{Lat}\}$ are as in (\ref{eq:x-ray-fe}), (\ref{eq:pe-ct}), and (\ref{eq:pe-xray}), respectively.
Figure 2 illustrates the architecture of proposed 3DPQT.

\subsection{Incorporating conditions into DDPM}
\label{sec:conditional diffusion model}

We extract 3D position-aware feature maps $\mathbf{c}^m_n$ in (\ref{eq:condition}) and now, we aim to incorporate its each element of size $1 \times 1 \times C_\mathcal{P}$ at its corresponding 2D spatial position.
The channel-wise concatenation is the most simple and widely-used conditioning method to accomplish this goal.
However, the channel-wise concatenation scheme did not fully utilize semantic information, leading to sub-optimal results \cite{spade,sdm}.
Therefore, we adopt SPADE \cite{spade} as our conditioning method. 
Following \cite{sdm}, we apply SPADE to our conditional DDPM.
We incorporate conditions $\mathbf{c}^m_{n}$ in (\ref{eq:condition}) into the feature maps extracted from an encoder and a decoder of denoising 2D U-Net $\mathcal{D}_{\theta_{\mathcal{D}}}$ with parameters $\theta_\mathcal{D}$ with the same 2D spatial resolution:
\begin{equation}
\label{eq:unet}
    \mathbf{\upepsilon}_{\theta} = \mathcal{D}_{\theta_{\mathcal{D}}}(\mathbf{x}_t,t,\mathbf{c}_n^m)
    ,~\text{where}~\theta = \{ \theta_\mathcal{E}, \theta_{\mathcal{P}},
    \theta_{\mathcal{Q}},
    \theta_\mathcal{T}, \theta_\mathcal{D} \},
\end{equation}
for $t \in [T]$, $n \in [N]$, and $ m \in \{\text{axial}, \text{coronal}, \text{sagittal}\}$.

We train DX2CT with conditions in an end-to-end fashion and define loss function as 
\begin{equation}
    \label{eq:final loss}
    \mathcal{L}{(\theta)} = \mathbb{E}_{\mathbf{x}_0,t,\mathbf{\upepsilon}, \mathbf{i}^{\text{All}},m,n} \left\| \mathbf{\upepsilon} - \mathbf{\upepsilon}_\theta(\mathbf{x}_t,t,\mathbf{i}^{\text{All}},m,n) \right\|_2^2,
\end{equation}
where $\theta$ is define as in (\ref{eq:unet}) and
$\mathbf{i}^{\text{All}} = \{\textbf{i}^{\text{PA}}, \textbf{i}^{\text{Lat}}\}$.

\begin{figure}[t!]
    \centering
    \includegraphics[width=0.9\linewidth]{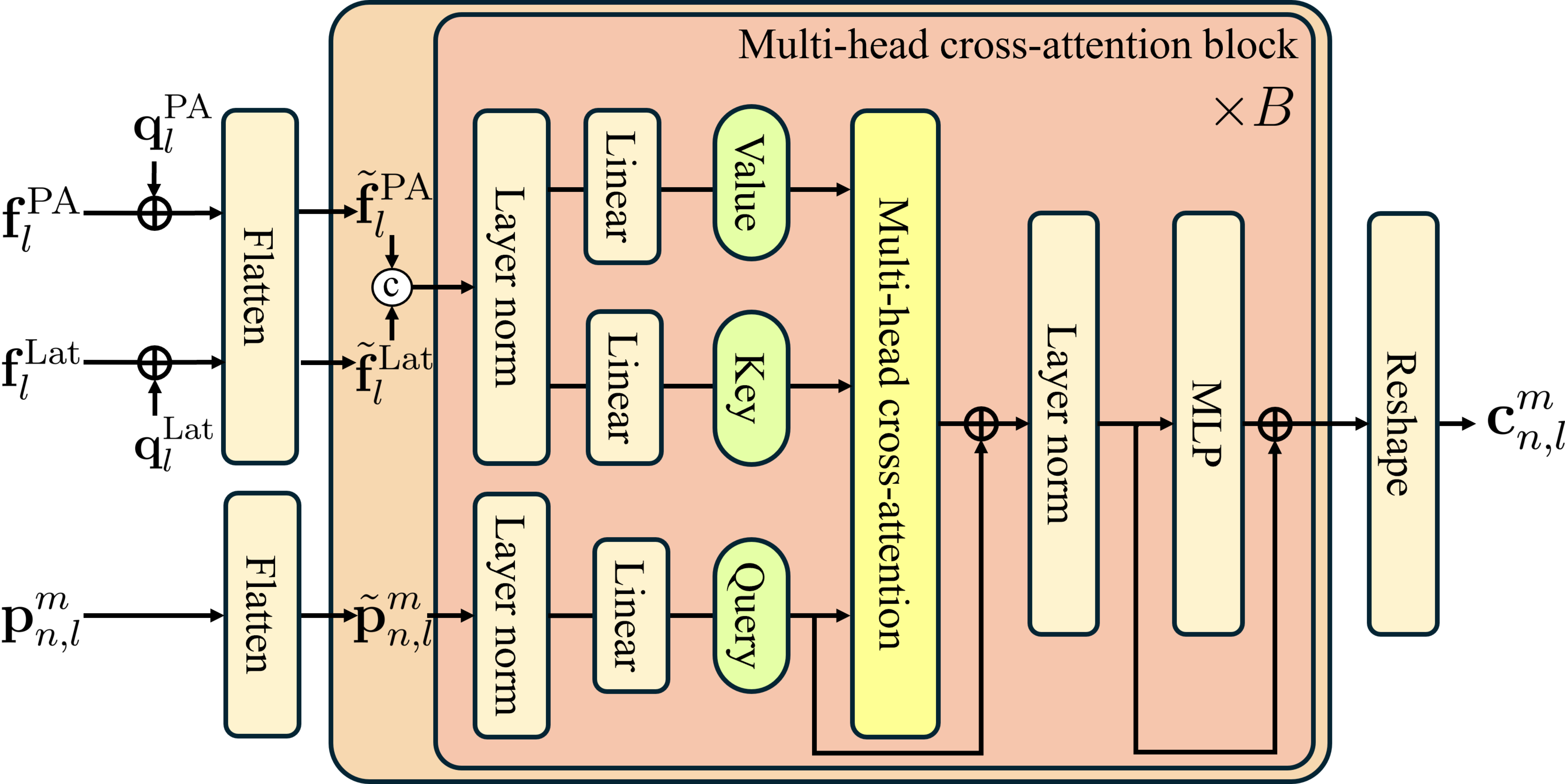}
    \caption{The architecture of proposed 3DPQT (at the $l$th scale).
    We use X-ray features $\textbf{f}^{\text{PA}}_l$ and $\textbf{f}^{\text{Lat}}_l$ in (\ref{eq:x-ray-fe}) for key and value.
    We use $\textbf{p}^{m}_{n,l}$ in (\ref{eq:pe-ct}) for query.
    We pass query, key, and value through $B$ multi-head cross-attention blocks and generate 3D position-aware feature map $\mathbf{c}^m_{n,l}$ in (\ref{eq:condition}).}
    \label{fig:3DPQT}
\end{figure}

\section{Results and discussion}
\label{sec: experiments}

This section describes the experimental setups and presents results with some discussion.
We compared proposed DX2CT with several SOTA 3D CT reconstruction methods using 2D X-ray(s).
In using biplanar X-rays, we compared DX2CT with PerX2CT \cite{perx2ct} and X2CTGAN \cite{x2ctgan};
in using monoplanar X-ray specifically the one in the PA view, we compared DX2CT with X2CTGAN \cite{x2ctgan} and 2DCNN \cite{single_tomo}.\footnote{
The runnable source codes and trained weights of \cite{diff2ct,difr3ct,xctdiff} are unavailable so we could not reproduce their results.}

\subsection{Experimental setups}

\noindent\textbf{Datasets and evaluation metrics.}
Similar to \cite{x2ctgan,perx2ct}, we used the benchmark LIDC \cite{lidc} CT dataset and synthesized X-rays using the digitally reconstructed radiographs \cite{drr}.
We preprocessed the synthesized dataset and sliced 3D CTs to 2D slices, following \cite{x2ctgan} and \cite{perx2ct}, respectively.
We divided $1,\!018$ pairs of a 3D CT volume and two 2D X-rays into $916$ training and $102$ test pairs.
To test trained models with real-world X-ray images, 
we followed the approach in \cite{x2ctgan,perx2ct}.
We transferred the style of synthetic X-ray images to real X-ray images using CycleGAN \cite{cyclegan} trained with the PadChest dataset \cite{padchest}.
As evaluation metrics, we used peak signal-to-noise ratio (PSNR), structural similarity index measure (SSIM) \cite{ssim}, and learned perceptual image patch similarity (LPIPS) \cite{lpips}.
We measured them with each of three reconstructed 3D CT volumes and averaged them.

\begin{figure}[t!]
\centering
\small\addtolength{\tabcolsep}{-3pt}
\renewcommand{\arraystretch}{0.3}
    \begin{tabular}
    {c@{\hspace{0.5mm}}c@{\hspace{1mm}}c@{\hspace{1mm}}c@{\hspace{1mm}}}
    \multicolumn{4}{c}{}
    \\
     {}&
     Axial &
     Coronal&
     Sagittal
     \\ 
     \raisebox{0.2\height}{\rotatebox{90}{{X2CTGAN}}} &
     \includegraphics[scale=0.6]{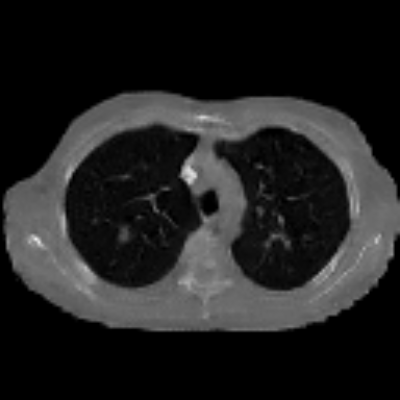} &
     \includegraphics[scale=0.6]{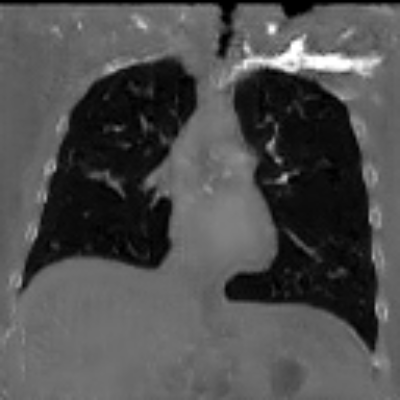} &
     \includegraphics[scale=0.6]{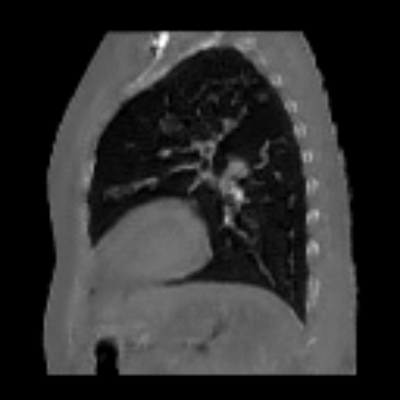} 
     \\
     \raisebox{0.1\height}{\rotatebox{90}{{$\text{PerX2CT}_\text{global}$}}} &
     \includegraphics[scale=0.6]{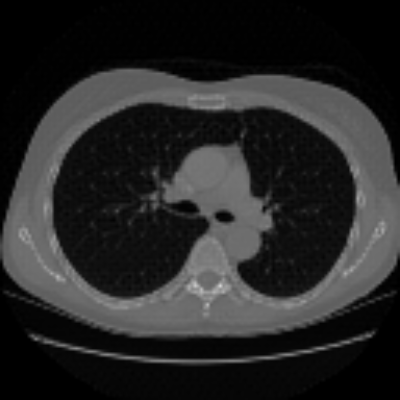} &
     \includegraphics[scale=0.6]{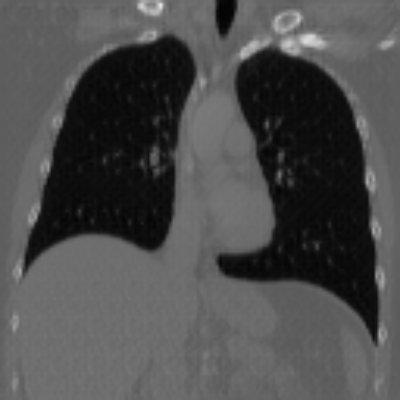} &
     \includegraphics[scale=0.6]{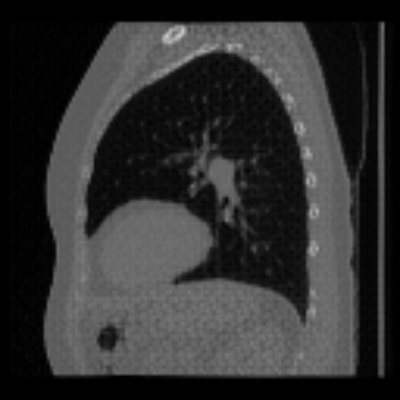} 
     \\   
     \raisebox{0.15\height}{\rotatebox{90}{{$\text{PerX2CT}_\text{local}$}}} &
     \includegraphics[scale=0.6]{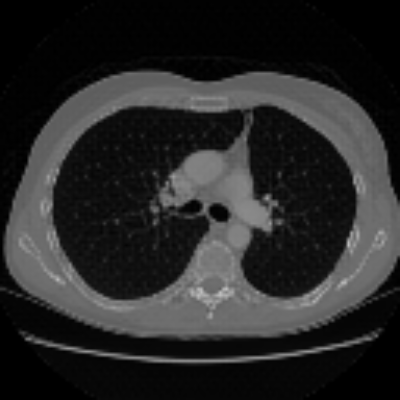} &
     \includegraphics[scale=0.6]{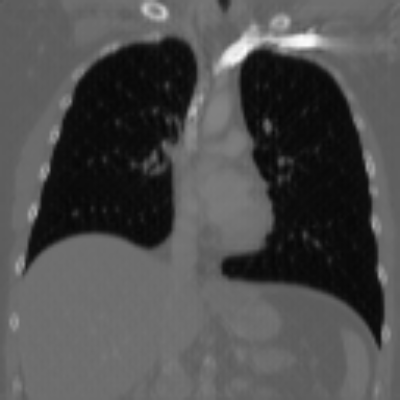} &
     \includegraphics[scale=0.6]{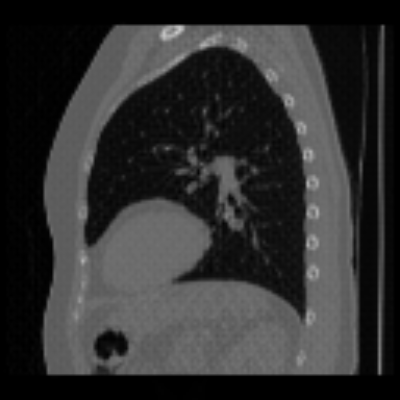} 
     \\
     \raisebox{0.50\height}{{\rotatebox{90}{\textbf{DX2CT}}}} &
     \includegraphics[scale=0.6]{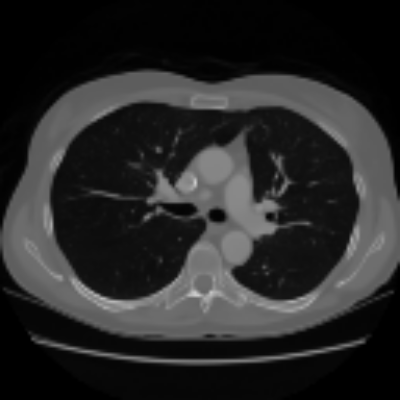} &
     \includegraphics[scale=0.6]{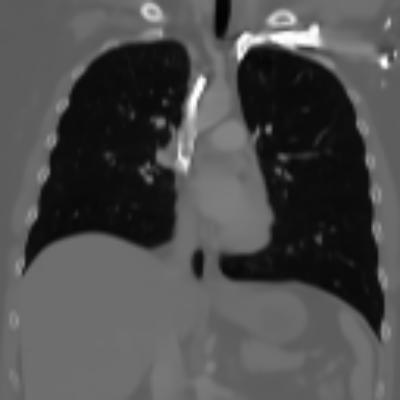} &
     \includegraphics[scale=0.6]{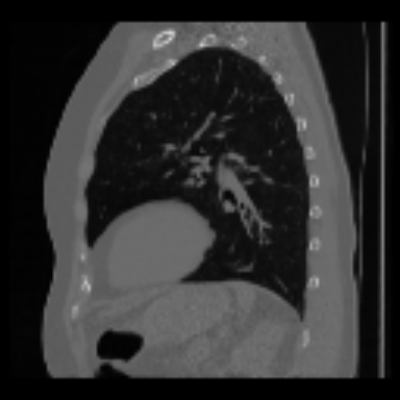} 
     \\
     \raisebox{0.1\height}{{\rotatebox{90}{{Ground-truth}}}} &
     \includegraphics[scale=0.6]{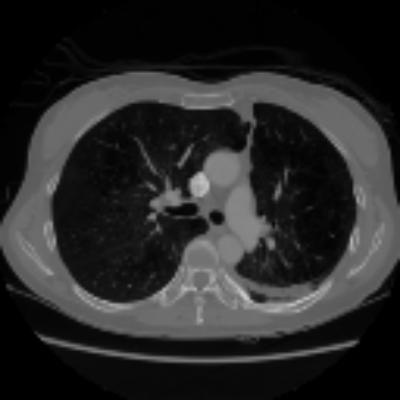} &
     \includegraphics[scale=0.6]{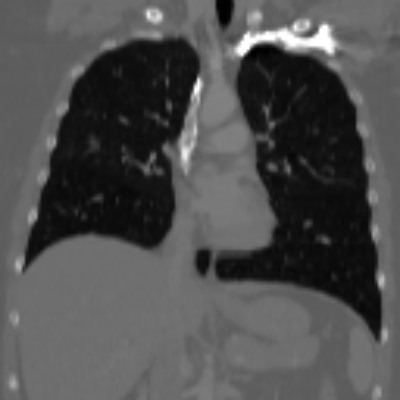} &
     \includegraphics[scale=0.6]{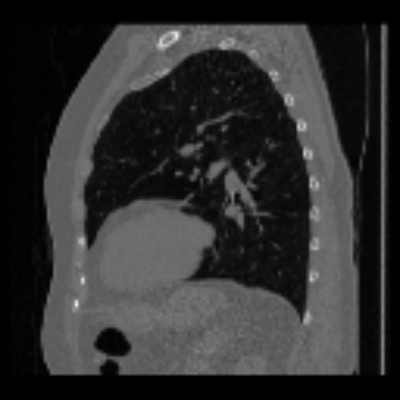}     
    \end{tabular}
    \caption{Comparisons of reconstructed 3D CTs with different methods (biplanar X-rays).}
    \label{fig:qualitative results}
\end{figure}

\noindent\textbf{Implementation details.}
As a feature extractor, we set $L$ as $3$, and used \textsf{conv2\_x}, \textsf{conv3\_x} and \textsf{conv4\_x} of ResNet-50 \cite{resnet} pre-trained with ImageNet \cite{imagenet}.
For a 3DPQT at each scale, we set $B$ as $12$.
For DX2CT, we used $T\!=\!1000$ and the linear noise schedule from $10^{-4}$ to $0.02$.
For denoising 2D U-Net of DX2CT \cite{2dunet}, we set the number of initial channels as $64$ and the channels multiple as $[1,1,2,3,4]$.
We trained DX2CT for $80$ epochs with a batch size of $16$ and used Adam optimizer with a learning rate of $5 \!\times\! 10^{-5}$.
For fast reconstruction, we used denoising diffusion implicit models (DDIM) \cite{ddim} with $50$ sampling steps.
To maintain reconstruction consistency among CT slices, we fixed the initial noise $\mathbf{x}_T$ and removed random noise term of DDIM.
In this way, reconstructed 2D slices in a 3D CT volume are only affected by incorporated conditions.
For existing SOTA methods, we used the default setups specified in their respective papers.
We used an NVIDIA GeForce RTX 4090 GPU throughout all the experiments.

\subsection{Comparisons between different 3D CT reconstruction models with 2D X-ray(s)}
\label{sec:comparison between other models}

Figure~\ref{fig:qualitative results} and Table~\ref{tab:numerical results}(a) using biplanar X-rays show that proposed DX2CT can outperform three existing SOTA methods.
Figure~\ref{fig:qualitative results} shows that DX2CT can provide more accurate overall shapes and details compared to the existing methods.
The quality of reconstructed CT slices in the axial plane is less satisfactory than those in the other planes. 
The reason is that the axial plane is perpendicular to the planes of biplanar X-rays so there exists less spatial (i.e., depth) information in the axial plane.
Without using the perceptual loss \cite{lpips}, proposed DX2CT gave comparable LPIPS results with PerX2CTs using \cite{lpips} in training. 
Compare their LPIPS results in Table~\ref{tab:numerical results}(a).

Table~\ref{tab:numerical results}(b) shows that DX2CT can achieve significantly better reconstruction performances compared to SOTA methods.

\begin{table}[t!]
    \centering
    \caption{Comparisons between different 3D CT reconstruction methods}
    \begin{tabular}{c | c | c | c}
    \specialrule{0.8pt}{1pt}{1pt}
    {} & PSNR ↑ & SSIM ↑ & LPIPS ↓ \\
    \hline
    \multicolumn{4}{c}{(a) From biplanar X-rays} \\ \hline
         X2CTGAN \cite{x2ctgan} & 26.747 & 0.647 & 0.316 \\
         $\text{PerX2CT}_{\text{global}}$ \cite{perx2ct} & 27.546 & 0.730 & 0.218 \\
         $\text{PerX2CT}_{\text{local}}$ \cite{perx2ct} & 27.659 & 0.739 & \textbf{0.210} \\
         \textbf{DX2CT (ours)} & \textbf{28.357} & \textbf{0.763} & 0.225 \\ \Xhline{1.7\arrayrulewidth}
    \multicolumn{4}{c}{(b) From monoplanar X-ray} \\ \hline
         2DCNN  \cite{single_tomo} & 24.471 & 0.549 & 0.427 \\
         X2CTGAN  \cite{x2ctgan} & 23.042 & 0.515 & 0.372 \\
         \textbf{DX2CT (ours)} & \textbf{25.506} & \textbf{0.643} & \textbf{0.277} \\
    \bottomrule
    \end{tabular}
    \label{tab:numerical results}
\end{table}

\subsection{Comparisons between different options in proposed DX2CT}
\label{sec:inject condition}

We evaluated the two key components of DX2CT, proposed 3DPQTs and the conditioning approach SPADE.
Comparing the results in the first and second rows with those in the third and fourth rows in Table~\ref{table:option} shows that the proposed 3DPQTs significantly improve the reconstruction performances by using 3D position-aware feature maps.
Comparing the results in the third and fourth rows in Table~\ref{table:option} demonstrates that SPADE is suitable for incorporating these modulated conditions.

\begin{table}[t!]
    \caption{Comparisons between different DX2CT variants}
    \label{table:option}
    \centering
    \begin{tabular}{c c | c c c}
    \toprule
        {3DPQTs} & {Conditioning} & {PSNR ↑} & {SSIM ↑} & {LPIPS ↓} \\
    \midrule
        X & Concat. & 26.939 & 0.721 & 0.251 \\
        X & SPADE  & 26.711 & 0.711 & 0.251 \\
        O & Concat. & 28.026 & 0.754 & 0.229 \\
        O & SPADE  & \textbf{28.357} & \textbf{0.763} & \textbf{0.225} \\
    \bottomrule
    \end{tabular}
\end{table}

\begin{figure}[t!]
\label{fig:real-world}
\centering
\small\addtolength{\tabcolsep}{-3pt}
\renewcommand{\arraystretch}{0.3}
\small
    \begin{tabular}{c@{\hspace{0mm}}c@{\hspace{2mm}}c@{\hspace{1mm}}c@{\hspace{1mm}}c@{\hspace{1mm}}c@{\hspace{1mm}}}
    \multicolumn{6}{c}{}
    \\
     {} &
     Real-style X-rays&
     {} &
     X2CTGAN &
     PerX2CT &
     \textbf{DX2CT}
     \\
     \raisebox{1.7\height}{\rotatebox{90}{PA}}&
     \includegraphics[scale=0.5]{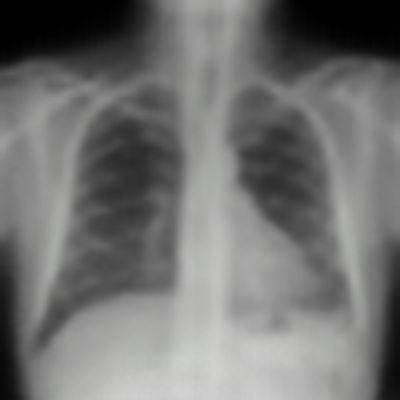} &
     \raisebox{0.3\height}{\rotatebox{90}{Coronal}} &
     \includegraphics[scale=0.5]{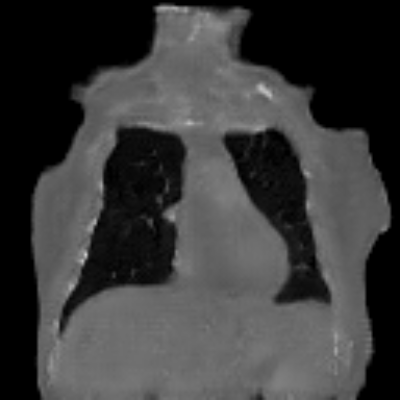} &
     \includegraphics[scale=0.5]{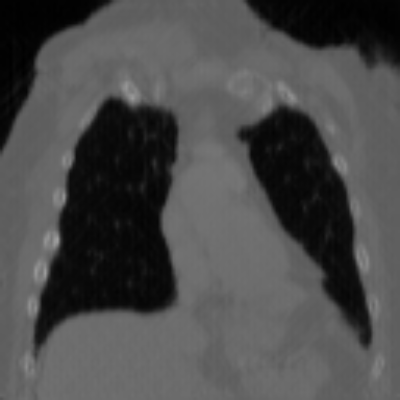} &
     \includegraphics[scale=0.5]{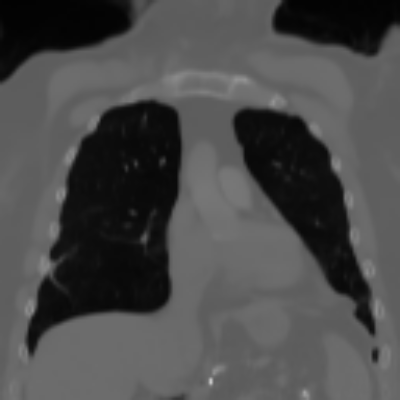} 
     \\
     \raisebox{0.45\height}{\rotatebox{90}{Lateral}}&
     \includegraphics[scale=0.5]{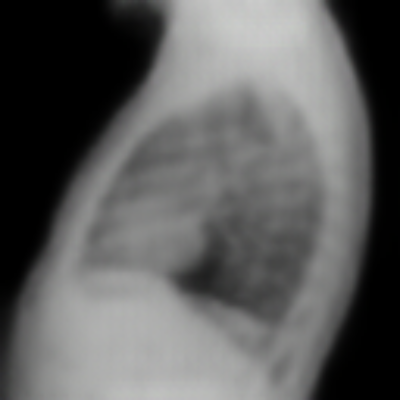} &
     \raisebox{0.4\height}{\rotatebox{90}{Sagittal}} &
     \includegraphics[scale=0.5]{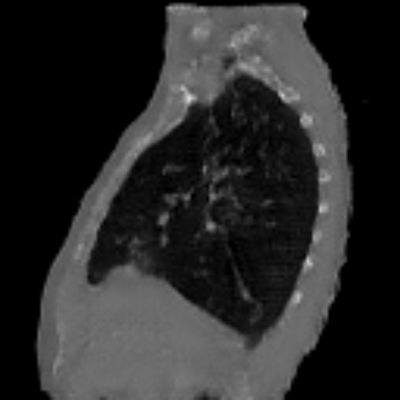} &
     \includegraphics[scale=0.5]{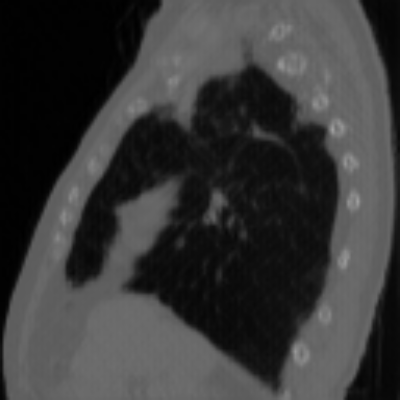} &
     \includegraphics[scale=0.5]{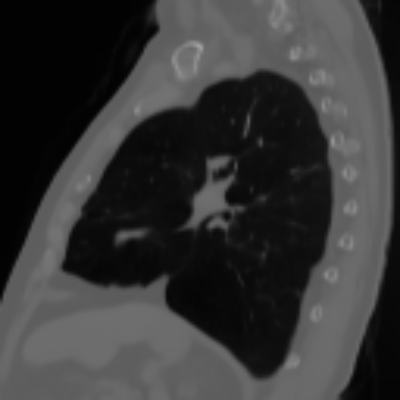} 
     \\
     {} &
     {} &
     \raisebox{0.8\height}{\rotatebox{90}{Axial}} &
     \includegraphics[scale=0.5]{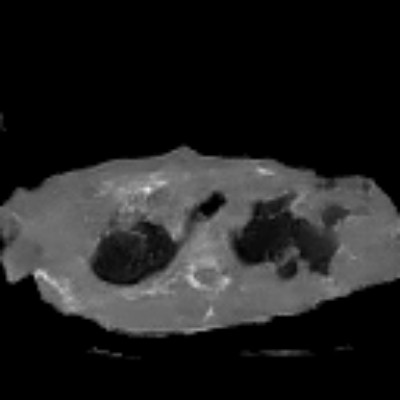} &
     \includegraphics[scale=0.5]{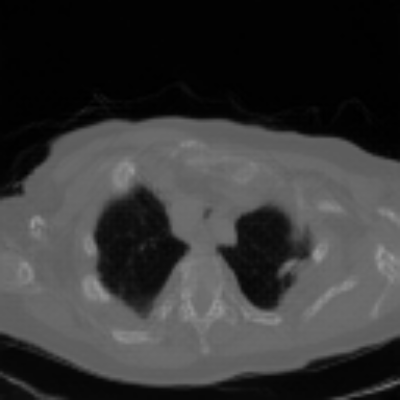} &
     \includegraphics[scale=0.5]{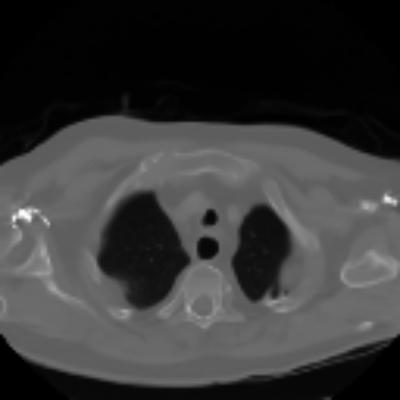} 
     
    \end{tabular}
    \caption{
    Comparisons of reconstructed 3D CTs with different methods (real-style biplanar X-rays).}
    \label{fig:real2sim}
\end{figure}

\subsection{Real-world data experiments}
\label{sec:real2sim}

The qualitative results in Figure~\ref{fig:real2sim} with real-world biplanar X-rays show that the structure of organs in reconstructed CTs by DX2CT better resembles to that of X-rays and DX2CT reconstructs sharper results, compared to the SOTA methods.

\section{Conclusion}
\label{sec: conclusion}

Reducing radiation exposure is important in medical imaging.
One can remarkably reduce radiations by reconstructing a 3D CT volume from only a few 2D X-rays, but this is a challenging task.
Our proposed DX2CT can take our steps toward to accomplish this task, by proposing a transformer that can provide conditions with rich information of target CTs to a conditional diffusion model.

\clearpage
\newpage

\bibliographystyle{IEEEtran}
\bibliography{ref}

\begin{thebibliography}{10}
\providecommand{\url}[1]{#1}
\csname url@samestyle\endcsname
\providecommand{\newblock}{\relax}
\providecommand{\bibinfo}[2]{#2}
\providecommand{\BIBentrySTDinterwordspacing}{\spaceskip=0pt\relax}
\providecommand{\BIBentryALTinterwordstretchfactor}{4}
\providecommand{\BIBentryALTinterwordspacing}{\spaceskip=\fontdimen2\font plus
\BIBentryALTinterwordstretchfactor\fontdimen3\font minus \fontdimen4\font\relax}
\providecommand{\BIBforeignlanguage}[2]{{%
\expandafter\ifx\csname l@#1\endcsname\relax
\typeout{** WARNING: IEEEtran.bst: No hyphenation pattern has been}%
\typeout{** loaded for the language `#1'. Using the pattern for}%
\typeout{** the default language instead.}%
\else
\language=\csname l@#1\endcsname
\fi
#2}}
\providecommand{\BIBdecl}{\relax}
\BIBdecl

\bibitem{single_tomo}
P.~Henzler, V.~Rasche, T.~Ropinski, and T.~Ritschel, ``Single-image tomography: 3d volumes from 2d cranial x-rays,'' \emph{Computer Graphics Forum}, vol.~37, no.~2, pp. 377--388, 2018.

\bibitem{single_proj}
L.~Shen, W.~Zhao, and L.~Xing, ``Patient-specific reconstruction of volumetric computed tomography images from a single projection view via deep learning,'' \emph{Nature biomedical engineering}, vol.~3, pp. 880--888, 2019.

\bibitem{x2ctgan}
X.~Ying, H.~Guo, K.~Ma, J.~Wu, Z.~Weng, and Y.~Zheng, ``X2ct-gan: Reconstructing ct from biplanar x-rays with generative adversarial networks,'' in \emph{Proc. {IEEE} Computer Vision and Pattern Recognition ({CVPR}'19)}, 2019, pp. 10\,611--10\,620.

\bibitem{ccxraynet}
M.~A.~R. Ratul, K.~Yuan, and W.~Lee, ``Ccx-raynet: A class conditioned convolutional neural network for biplanar x-rays to ct volume,'' in \emph{Proc. {IEEE} International Symposium on Biomedical Imaging ({ISBI}'21)}, 2021, pp. 1655--1659.

\bibitem{cvae-gan}
L.~Jiang, M.~Zhang, R.~Wei, B.~Liu, X.~Bai, and F.~Zhou, ``Reconstruction of 3d ct from a single x-ray projection view using cvae-gan,'' in \emph{Proc. {IEEE} International Conference on Medical Imaging Physics and Engineering ({ICMIPE}'21)}, 2021, pp. 1--6.

\bibitem{x-ctrsnet}
R.~Ge \emph{et~al.}, ``X-ctrsnet: 3d cervical vertebra ct reconstruction and segmentation directly from 2d x-ray images,'' \emph{Knowledge-Based System}, vol. 236, p. 107680, 2022.

\bibitem{perx2ct}
D.~Kyung, K.~Jo, J.~Choo, J.~Lee, and E.~Choi, ``Perspective projection-based 3d ct reconstruction from biplanar x-rays,'' in \emph{Proc. {IEEE} International Conference on Acoustics, Speech and Signal Processing ({ICASSP}'23)}, 2023, pp. 1--5.

\bibitem{xctdiff}
\BIBentryALTinterwordspacing
Q.~Bai, T.~Liu, Z.~Liu, D.~T. Y.~Tong, and J.~Udupa, ``Xctdiff: Reconstruction of ct images with consistent anatomical structures from a single radiographic projection image,'' 2024, unpublished. [Online]. Available: \url{arXiv:2406.04679}
\BIBentrySTDinterwordspacing

\bibitem{diff2ct}
\BIBentryALTinterwordspacing
Q.~Zhi \emph{et~al.}, ``Diff2ct: Diffusion learning to reconstruct spine ct from biplanar x-rays,'' 2024, unpublished. [Online]. Available: \url{arXiv:2408.09731}
\BIBentrySTDinterwordspacing

\bibitem{difr3ct}
\BIBentryALTinterwordspacing
Y.~Sun \emph{et~al.}, ``Difr3ct: Latent diffusion for probabilistic 3d ct reconstruction from few planar x-rays,'' 2024, unpublished. [Online]. Available: \url{arXiv:2408.15118}
\BIBentrySTDinterwordspacing

\bibitem{gan}
I.~Goodfellow \emph{et~al.}, ``Generative adversarial nets,'' in \emph{Advances in Neural Information Processing Systems ({NeurIPS}'14)}, vol.~27, 2014.

\bibitem{ddpm}
J.~Ho, A.~Jain, and P.~Abbeel, ``Denoising diffusion probabilistic models,'' in \emph{Proc. Advances in Neural Information Processing Systems ({NeurIPS}'20)}, vol.~33, 2020.

\bibitem{ddim}
J.~Song, C.~Meng, and S.~Ermon, ``Denoising diffusion implicit models,'' in \emph{Proc. International Conference on Learning Representations ({ICLR}'21)}, 2021.

\bibitem{ldm}
R.~Rombach, A.~Blattmann, D.~Lorenz, P.~Esser, , and B.~Ommer, ``High-resolution image synthesis with latent diffusion models,'' in \emph{Proc. {IEEE} Computer Vision and Pattern Recognition ({CVPR}'22)}, June 2022, pp. 10\,684--10\,695.

\bibitem{sr3}
C.~Saharia, J.~Ho, W.~Chan, T.~Salimans, D.~J. Fleet, and M.~Norouzi, ``Image super-resolution via iterative refinement,'' \emph{{IEEE} Transactions on Pattern Analysis and Machine Intelligence}, vol.~45, no.~4, pp. 4713--4726, 2023.

\bibitem{sdm}
\BIBentryALTinterwordspacing
W.~Wang \emph{et~al.}, ``Semantic image synthesis via diffusion models,'' 2022, unpublished. [Online]. Available: \url{arXiv:2207.00050v2}
\BIBentrySTDinterwordspacing

\bibitem{ddrm}
B.~Kawar, M.~Elad, S.~Ermon, and J.~Song, ``Denoising diffusion restoration models,'' in \emph{Proc. Advances in Neural Information Processing Systems ({NeurIPS}'22)}, 2022, pp. 23\,593--23\,606.

\bibitem{light-field}
H.~S. Yun and I.~Y. Chun, ``Improving light field reconstruction from limited focal stack using diffusion models,'' in \emph{Proc. {IEEE} Machine Learning for Signal Processing ({MLSP}'24)}, 2024, accepted, to appear.

\bibitem{3dunet}
{\"O}.~\c{C}i\c{c}ek, A.~Abdulkadir, S.~S. Lienkamp, T.~Brox, and O.~Ronneberger, ``3d u-net: Learning dense volumetric segmentation from sparse annotation,'' in \emph{Proc. Medical Image Computing and Computer-Assisted Intervention ({MICCAI}'16)}, 2016, pp. 424--432.

\bibitem{vit}
A.~Dosovitskiy \emph{et~al.}, ``An image is worth 16x16 words: Transformers for image recognition at scale,'' in \emph{Proc. International Conference on Learning Representations ({ICLR}'21)}, 2021.

\bibitem{2dunet}
O.~Ronneberger, P.~Fischer, and T.~Brox, ``U-net: Convolutional networks for biomedical image segmentation,'' in \emph{Proc. Medical Image Computing and Computer-Assisted Intervention ({MICCAI}'15)}, 2015, pp. 234--241.

\bibitem{lidc}
S.~G.~A. III \emph{et~al.}, ``The lung image database consortium (lidc) and image database resource initiative (idri): A completed reference database of lung nodules on ct scans,'' \emph{Medical Physics}, vol.~38, no.~2, pp. 915--931, 2011.

\bibitem{transformer}
A.~Vaswani \emph{et~al.}, ``Attention is all you need,'' in \emph{Proc. Advances in Neural Information Processing Systems ({NeurIPS}'17)}, vol.~30, 2017.

\bibitem{pvt}
W.~Wang \emph{et~al.}, ``Pyramid vision transformer: A versatile backbone for dense prediction without convolutions,'' in \emph{Proc. {IEEE} International Conference on Computer Vision ({ICCV}'21)}, 2021, pp. 568--578.

\bibitem{dit}
W.~Peebles and S.~Xie, ``Scalable diffusion models with transformers,'' in \emph{Proc. {IEE} International Conference on Computer Vision ({ICCV}'23)}, October 2023, pp. 4195--4205.

\bibitem{detr}
N.~Carion, F.~Massa, G.~Synnaeve, N.~Usunier, A.~Kirillov, and S.~Zagoruyko, ``End-to-end object detection with transformers,'' in \emph{Proc. {IEEE} European Conference on Computer Vision ({ECCV}'20)}, 2020, pp. 213--229.

\bibitem{nerf}
B.~Mildenhall, P.~P. Srinivasan, M.~Tancik, J.~T. Barron, R.~Ramamoorthi, and R.~Ng, ``Nerf: Representing scenes as neural radiance fields for view synthesis,'' in \emph{Proc. The European Conference on Computer Vision ({ECCV}'20)}, 2020, pp. 405--421.

\bibitem{spade}
T.~Park, M.-Y. Liu, T.-C. Wang, and J.-Y. Zhu, ``Semantic image synthesis with spatially-adaptive normalization,'' in \emph{Proc. {IEEE} Computer Vision and Pattern Recognition ({CVPR}'19)}, 2019, pp. 2322--2341.

\bibitem{drr}
N.~Milickovic, D.~Baltas, S.~Giannouli, M.~Lahanas, and N.~Zamboglou, ``Ct imaging based digitally reconstructed radiographs and their application in brachytherapy,'' \emph{Physics in Medicine \& Biology}, vol.~45, no.~10, p. 2787, 2000.

\bibitem{cyclegan}
J.-Y. Zhu, T.~Park, P.~Isola, , and A.~A. Efros, ``Unpaired image-to-image translation using cycle-consistent adversarial networks,'' in \emph{Proc. {IEEE} International Conference on Computer Vision ({ICCV}'17)}, 2017, pp. 2242--2251.

\bibitem{padchest}
A.~Bustos, A.~Pertusa, J.-M. Salinas, and M.~de~la Iglesia-Vayá, ``Padchest: A large chest x-ray image dataset with multi-label annotated reports,'' \emph{Medical Image Analysis}, vol.~66, p. 101797, 2020.

\bibitem{ssim}
Z.~Wang, A.~C. Bovik, H.~R. Sheikh, , and E.~P. Simoncelli, ``Image quality assessment: from error visibility to structural similarity,'' \emph{{IEEE} Transactions on Image Processing}, vol.~13, no.~4, pp. 600--612, 2004.

\bibitem{lpips}
R.~Zhang, P.~Isola, A.~A. Efros, E.~Shechtman, and O.~Wang, ``The unreasonable effectiveness of deep features as a perceptual metric,'' in \emph{Proc. {IEEE} Computer Vision and Pattern Recognition ({CVPR}'18)}, 2018, pp. 586--595.

\bibitem{resnet}
K.~He, X.~Zhang, S.~Ren, and J.~Sun, ``Deep residual learning for image recognition,'' in \emph{Proc. {IEEE} Computer Vision and Pattern Recognition ({CVPR}'16)}, 2016, pp. 770--778.

\bibitem{imagenet}
J.~Deng, W.~Dong, R.~Socher, L.-J. Li, K.~Li, and L.~Fei-Fei, ``Imagenet: A large-scale hierarchical image database,'' in \emph{Proc. {IEEE} Computer Vision and Pattern Recognition ({CVPR}'09')}, 2019, pp. 248--255.

\end{thebibliography}

\end{document}